\documentclass[aps,pra,superscriptaddress,balancelastpage,twocolumn]{revtex4}
\usepackage{eurosym}
\usepackage{bbding}
\usepackage{amsthm}
\usepackage{dcolumn}
\usepackage{bm}
\usepackage{subfigure}
\usepackage{epsfig,graphicx,times}
\usepackage{amstext}
\usepackage{amsmath}
\usepackage{amssymb}
\usepackage{graphicx}
\usepackage{latexsym}
\usepackage{bm}
\usepackage[colorlinks,citecolor=blue, linkcolor=blue,hyperindex,dvipdfm]{hyperref}
\usepackage{xcolor}
\usepackage[utf8]{inputenc}
\begin{document}

\title{Controllable Fano-type optical response and four-wave mixing via
magnetoelastic coupling in a opto-magnomechanical system}
\author{Amjad Sohail}
\email{amjadsohail@gcuf.edu.pk}
\affiliation{Department of Physics, Government College University, Allama Iqbal Road,
Faisalabad 38000, Pakistan.}
\author{Rizwan Ahmed}
\affiliation{Physics division, Pakistan institute of nuclear science and technology
(PINSTECH), Nilore, Islamabad 45650 Pakistan}
\author{Jia-Xin Peng}
\email{18217696127@163.com}
\affiliation{State Key Laboratory of Precision Spectroscopy, Quantum Institute for Light
and Atoms, Department of Physics, East China Normal University, Shanghai
200062, People's Republic of China}
\author{Aamir Shahzad}
\affiliation{Department of Physics, Government College University, Allama Iqbal Road,
Faisalabad 38000, Pakistan.}
\author{Tariq Munir}
\affiliation{Department of Physics, Government College University, Allama Iqbal Road,
Faisalabad 38000, Pakistan.}
\author{S. K. Singh}
\email{ singhshailendra3@gmail.com}
\affiliation{Graphene and Advanced 2D Materials Research Group (GAMRG), School of
Engineering and Technology, Sunway University, No. 5, Jalan Universiti,
Bandar Sunway, 47500 Petaling Jaya, Selangor, Malaysia}
\author{Marcos C\'{e}sar de Oliveira}
\email{ marcos@ifi.unicamp.br}
\affiliation{Instituto de F\'{\i}sica Gleb Wataghin, Universidade Estadual de Campinas,
Campinas, SP, Brazil}

\begin{abstract}
We analytically investigate the Fano-type optical response and four-wave mixing (FWM) process by exploiting the magnetoelasticity of a ferromagnetic material. The deformation of the ferromagnetic material plays the role of mechanical displacement, which is simultaneously coupled to both optical and magnon modes. We report that the
magnetostrictively induced displacement demonstrates Fano profiles, in the output field, which is well-tuned by adjusting the system parameters, like effective magnomechanical coupling, magnon detuning, and cavity detuning. It is found that the magnetoelastic interaction also gives rise to the FWM phenomenon. The number of the FWM
signals mainly depends upon the effective magnomechanical coupling and the magnon detuning. Moreover, the FWM spectrum exhibits suppressive behavior upon increasing (decreasing) the magnon (cavity) decay rate. The present scheme will open new perspectives in highly sensitive detection and quantum information processing.
\end{abstract}

\maketitle




\section{Introduction}
Cavity optomechanics is a rapidly growing area of theoretical and
experimental research that explores the coherent coupling between mechanical
and optical modes through the radiation pressure force of photons being
trapped inside optomechanical cavities \cite{Mey}. It has been responsible
for significant advances in quantum technology, particularly when
non-linearities are available. We mention quantum information processing
\cite{Fio}, quantum walk implementation \cite{Jalil}, ultrahigh-precision
measurement\cite{Teu,AliM}, higher-order sidebands generation \cite{Qian},
optical nonlinearity \cite{Lud}, single photon blockade \cite{SS1,SS2},
quantum entanglement \cite{AM0,MYP,SKS,BTT}, gravitation-wave detection\cite%
{Arva,AliM2}, optomechanically induced transparency (OMIT) \cite{MYPOM1,OM1,OM2}, and
optomechanically induced absorption (OMIA) \cite{MYPOM2} including
optomechanically induced stochastic resonance \cite{Faraz}. In OMIT, analogously to electromagnetic
induced transparency (EIT) \cite{Imam}, the radiation pressure-induced
mechanical oscillations lead to destructive interference and hence
significantly alter the optical response as well as its group delay \cite%
{Saif,Saiff}.

Another important and fascinating non-linear phenomenon is the interference and quantum coherence, called Fano resonance \cite{Yuri}, where the prominent feature is a steep and sharp asymmetric line profile. Fano resonance were theoretically discovered by Ugo Fano \cite{Fano}, who found that the interaction of a discrete excited state of an
atom with a continuum of scattering states, is entirely different from the resonance phenomena delineated by the Lorentzian formula \cite{Fano}. Fano interference has important applications such as in lasing without population inversion \cite{Gav}, as a probe to study decoherence \cite{Kuh}, and as a way to improve the efficiency of heat engines \cite{Chap}. Moreover, the
radiating two coupled-resonators model, in which the two resonators have quite different lifetime can be used to explain the Fano line shapes \cite{Tass,Alim2}. So the lifetimes of these modes are very significant in observing the Fano line shapes \cite{Gal} and can be easily obtained in plasmonic systems \cite{Gall} and metamaterial structures \cite{Ali,AliM5}. Fano resonance has been recently studied in cavity magnomechanics \cite{Kamran} and magnon-qubit system \cite{Sabur}.

Four-wave mixing (FWM), as the most engrossing nonlinear optical effect, is widely studied in optomechanical systems \cite{FWM1}. FWM is based on interference and quantum coherence and is extensively used to enlarge the frequency span of coherent light sources to ultraviolet and infrared \cite{FWM2,AliM3}. Multi-wave or FWM-mixing processes allow many vital physical processes, such as realizing optical bistability \cite{FWM3}, normal mode splitting \cite{FWM4A} or generating two coexisting pairs of narrowband biphotons \cite{FWM5}. Consequently, the FWM effect has gained much attention from researchers. Jiang et al. examined the controllability and tunability of the FWM process by driving the two cavities in a two-mode cavity optomechanical system at their respective red sidebands \cite{FWM6}. Wang and Chen studied the enhancement of the FWM response \cite{FWM7}. It
must be pointed out that FWM is a weak phenomenon, which makes it hard to be realized in experiments. Therefore new directions are being proposed to improve the strength of those nonlinear phenomena.

The hybridization of different modes in cavity magnomechancis (CMM) opens up a promising platform to explore numerous non-linear quantum effects \cite{Tang}. Magnomechanical systems consider a geometry such that an optical mode directly (indirectly) couples with a magnon mode (mechanical mode).
They are based totally on magnons-quanta of spin waves in Ferrimagnetic materials \cite{Heyroth}, such as yttrium iron garnet (YIG), which provide a versatile and new platform to investigate strong light-matter interactions \cite{Gros}. Strong interactions are due to two main reasons: (1) The Spin density of the YIG material is very high; (2) The damping rate of the YIG
material is very low. Therefore, in the last decade, CMM has been explored in theoretical as well as an experimental domain because of its potential applications for observing macroscopic quantum states \cite{Jiee}, quantum sensing \cite{Usami} and magnomechanically induced transparency (MMIT) \cite{Kamran,AliM4,XLi}. In CMM systems, the microwave (MW) cavity field is coupled to a magnon mode in a ferromagnetic YIG sphere \cite{MYP} as well as the magnon mode also couples to a (vibrational) mechanical mode of the sphere through the magnetostrictive force \cite{Deng}. Such magnomechanical interaction is
dispersive \cite{Tang,Rome} in a similar fashion to radiation pressure interaction, leading to several theoretical and experimental investigations to study various quantum phenomena in cavity magnomechanical systems analogous to cavity optomechanics, such as magnomechanically induced transparency (MMIT) \cite{Tang,Kamran}, magnetically tunable slow light \cite%
{Cui,XLi}, exceptional points \cite{Den}, tripartite magnon-photon-phonon entanglement and Einstein-Podolsky-Rosen steerable nonlocality \cite{Tann}, cooling of mechanical motion \cite{Asjad}, phonon lasing \cite{Cong} and parity-time-related phenomena \cite{Annl}.

In this paper, we propose an unconventional and advantageous
opto-magnomechanical system (OMMS) setup which is a composite architecture
based on the indirect coupling of optical photons and magnon modes via the
phononic mode (vibrational mode of the magnon) as shown in Fig. 1(a). The
magnetoelastic coupling is responsible for the geometric deformation of the
magnon in a YIG bridge which forms the vibrational motion. We consider that
the phonon frequency is much smaller than the frequency of the magnon and in
this situation, the dispersive interaction between the phonon and magnon
modes becomes dominant (See Appendix A for details). This present scheme can
be well fabricated by placing a high reflectivity small size mirror on the
surface (edge) of the YIG bridge and therefore, the vibrational motion of
the magnon can be easily coupled to an optical cavity through the radiation
pressure, assembling an optomechanical cavity. However, the tiny size of the
mirror does not affect the vibrations of the YIG bridge. In addition, such
kind of indirect coupling between optical photons and magnons can easily be
realized by assuming the low mechanical damping rate $\gamma_{b}$ both in
the optomechanical and magnomechanical cooperativities $C_{om}=\frac{%
G_{c}^{2}}{\kappa_{c}\gamma_{b}}>1$ ($\kappa_{c}$ being the cavity decay
rate), and $C_{mm}=\frac{G_{m}^{2}}{\kappa_{m}\gamma_{b}}>1$ ($\kappa_{m}$
being the magnon decay rate) respectively, under present-day technology \cite%
{Mey,Tang,CAPE}. Moreover, experimentally feasible parameters have been
taken for the current scheme. In this paper, we have presented several
interesting results for generating/controlling Fano profiles and the FWM by
varying many system parameters.

The current article is arranged as follows. In Sect. 2, we present the
theoretical model and the corresponding Hamiltonian of the OMMS. Section 3
designates the dynamics of the OMMS by exploiting the standard quantum
Langevin approach, to get the expression for the optical response and the FWM
process. In Sect. 4, we analytically discuss the optical response and the
FWM process in detail. Finally, the conclusion is given in the last Sect. 5.
\section{The model and the dynamics}
The opto-magnomechanical system under consideration consists of an optical
mode $c$ and a slab of YIG nanoresonator in which collective excitation of a
number of spins are coupled to mechanical vibrations \cite{Heyroth}. It is
vital to mention here that a micrometer size YIG bridge can well support a
gigahertz (megahertz) magnon mode (vibrational mode) \cite{Heyroth}.
Therefore, such a magnomechanical system has a strong magnon-phonon
dispersive coupling \cite{disp2,mag,Kittel}. The present scheme can be
accomplished by placing a small mirror on the edge (surface) of the YIG
bridge as shown in Fig. 1 (a), and in this way, an optical field can be coupled
to the magnetostriction-induced mechanical displacement via radiation
pressure force. The fabricated mirror should be highly refractive, light,
and small such that it will not affect the mechanical
properties of the ferrimagnet. Therefore, in this way the
magnomechanical displacement can be probed by light
exploiting the optomechanical interaction, which is further
used to determine the magnon population \cite{ZRCS}. Moreover, the deformation mode (displacement of the mechanical
resonator) is along the direction of the sticky surface area of the slab,
such that the tightly coupled YIG bridge and the mirror resonate with the
same frequency. In addition, the dispersive magnetostrictive interaction
between magnon and phonon modes depends on the resonance frequencies of
these coupled modes \cite{Zha}.

It is important to mention here that Fig.1 is only a diagrammatic sketch of one of the possible
configurations to couple the magnomechanically induced displacement to an optical cavity. In addition, one doesn't have to attach a mirror (on the surface of the microbridge) to form the optical cavity. Alternatively, one can also adopt the analogy of taking a "membrane-in-the-middle" configuration by positioning the YIG sample inside the optical cavity. In this setup, the  magnomechanically induced displacement can also be dispersively coupled to the optical cavity. In this case, no additional damping of the mechanical mode is needed. In the current study, we consider a large size crystal such that the magnon frequency is considered to be much greater than the vibrational frequency of the phononic mode such that the dispersive interaction between magnon and phonon modes, becomes dominant \cite{Tang,Ballestero}. The Hamiltonian of the present OMMS reads
\begin{equation}
H/\hbar =H_{\circ }+H_{in}+H_{dr},
\end{equation}%
where
\begin{eqnarray}
H_{\circ } &=&\omega _{c}c^{\dagger }c+\omega _{m}m^{\dagger }m+\frac{\omega
_{b}}{2}(q^{2}+p^{2}), \\
H_{in} &=&g_{mb}m^{\dagger }mq-g_{cb}c^{\dagger }cq,  \label{Hi} \\
H_{dr} &=&i\Omega (m^{\dagger }e^{-i\omega _{\circ }t}-me^{i\omega _{\circ
}t})  \notag \\
&&+i\sum_{f=L,p} \varepsilon _{f}(c^{\dagger }e^{-i\omega _{f}t}-ce^{i\omega
_{f}t}) .
\end{eqnarray}%
\begin{figure}[b!]
\par
\begin{center}
\includegraphics[width=0.85\columnwidth,height=3.5in]{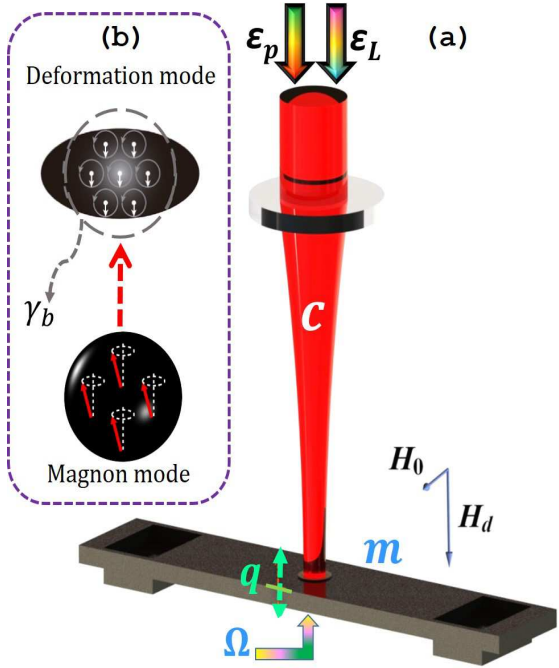}
\end{center}
\caption{(Color Online) (a) Schematic diagram of the opto-magnomechanical
system where mechanical displacement is simultaneously coupled to the
magnonic mode via the dispersive magnetostrictive interaction and to an
optical cavity through radiation pressure force. (b) Schematic diagram of
the Kittel and deformation modes.}
\label{A}
\end{figure}
where $c(m)$ and $c^{\dag }\left( m^{\dag }\right) $ are the respective
annihilation and creation operators for the cavity (magnon) mode. $p$ and $q
$ are the dimensionless momentum and position of the phononic mode
(deformation vibrational mode), considered as a mechanical resonator, which
simultaneously couples to the cavity (magnon) mode via radiation pressure
interaction $g_{cb}$ (dispersive magnetostrictive interaction $g_{mb}$). The
details of the optomechanical coupling $g_{cb}$ can be seen in \cite{CGEN}
while the detailed discussion about the magnomechanical coupling $g_{mb}$ is
present in Appendix A. In addition, $\omega _{c}$($\omega _{m}$)[$\omega_{b}$%
] is the resonance frequencies of the cavity mode (magnon mode) [phononic
mode]. $\omega _{m}$ can be flexibly tuned via $\omega _{m}=\gamma _{0}H_{m}$%
, where $H_{m}$ is the bias magnetic field and $\gamma _{0}$ is the
gyromagnetic ratio. The Rabi frequency $\Omega =\left( \sqrt{5}/4\right)
\gamma _{0}\sqrt{N_{s}}B_{0} $ \cite{Jiee} represents the coupling strength
of the input laser drive field with frequency $\omega _{0}$, where $\gamma
_{0}=28$GHz/T, $B_{0}=3.9\times 10^{-9}$T and the total number of spins $%
N_{s}=\rho V$, where $\rho =4.22\times 10^{27}m^{-3}$ denotes the spin
density and $V$ being the volume of the YIG bridge. In addition, we apply
the Holstein-Primakoff transformation to the bosonic operators for the
magnon ($m$ and $m^{\dagger }$), which basically form from the collective
motion of the spins. In fact, $\Omega $ can be derived by the basic
low-lying excitations assumption i.e., $\langle m^{\dagger }m\rangle \ll 2Ns
$, where $N$ ($s=\frac{5}{2}$) is the total number of spin states (the spin
number of the ground state Fe$^{3+}$ ion) in YIG \cite{Holstein}.
Furthermore, we also drive the optical cavity through external laser field
with amplitude $\varepsilon_{L} =\sqrt{\kappa _{c}\wp _{L}/\hbar \omega _{L}}
$, where $\kappa _{c}$ is the decay rate of the optical cavity mode, $\omega
_{L}$ ($\wp _{L}$) is the frequency (power) of the external input laser
field. At the same time, the cavity mode is also driven by a weak probe
field with amplitude $\varepsilon_{L}$ and frequency $\wp _{p}$. The
corresponding Hamiltonian of the OMMS, after the rotating wave approximation
at the coupling frequency $\omega _{L}$, is given by
\begin{eqnarray}
H/\hbar &=&\Delta _{c}^{0}c^{\dag }c+\Delta _{m}^{0}m^{\dag }m+\frac{\omega
_{b}}{2}(q^{2}+p^{2})+g_{mb}m^{\dag }mq  \notag \\
&&-g_{cb}c^{\dagger }cq+i\Omega (m^{\dag }-m)+i\varepsilon _{L}(c^{\dagger
}-c)  \notag \\
&&+i\varepsilon _{p}(c^{\dagger }e^{-i\delta t}-ce^{i\delta t}),
\end{eqnarray}%
where $\Delta _{m}^{0}=\omega _{m}-\omega _{L}$, $\Delta _{c}^{0}=\omega
_{c}-\omega _{L}$ and $\delta =\omega _{p}-\omega _{L}$.

Now, we discuss the dynamics of this OMMS by writing the corresponding QLEs,
which are given by
\begin{eqnarray}
\dot{q} &=&\omega _{b}p,  \notag \\
\dot{p} &=&\omega _{b}q-\gamma _{b}p-g_{mb}m^{\dagger }m+g_{cb}c^{\dagger
}c+\xi , \notag \\
\dot{c} &=&-(i\Delta _{c}^{0}+\kappa _{c})c+ig_{cb}cq+\varepsilon _{L}
+\varepsilon _{p}e^{-i\delta t}+\sqrt{\kappa _{c}}c^{in}, \notag \\
\dot{m} &=&-(i\Delta _{m}^{0}+\kappa _{m})m-ig_{mb}mq+\Omega +\sqrt{\kappa
_{m}}m^{in},  \label{M}
\end{eqnarray}%
where $\kappa _{c}$($\kappa _{m}$) is the decay rate of the cavity mode
(magnon mode), and $\gamma _{b}$ is the damping rate of the vibrational
mode. Furthermore, $\xi $, $c^{in}$ and $m^{in}$ are, respectively, the
input noise operators for the vibrational mode, cavity modes and magnon
mode. The above QLEs can be linearized by writing $z=z_{s}+\delta z$, $%
(z=p,q,c,m)$, where $z_{s}$ ($\delta z$) is the steady state value
(fluctuation) of any operator. For a sufficiently large amplitude of magnon
and cavity modes, average value should be much greater than the fluctuation.
Substituting the above ansatz into Eq.(\ref{M}) and ignoring the
higher order perturbations, the average values of the dynamical operators
are given by
\begin{eqnarray}
p_{s} &=&0, \notag \\
q_{s} &=&g_{cb}\frac{\left\vert c_{s} \right\vert ^{2}%
}{\omega _{b}}-g_{cb}\frac{\left\vert m_{s} \right\vert ^{2}}{\omega _{b}},  \notag \\%
c_{s} &=&\frac{\varepsilon _{L}}{(i\Delta _{c}+\kappa _{c})}, \notag \\
m_{s} &=&\frac{\Omega }{(i\Delta _{m}+\kappa _{m})}, \label{SS}
\end{eqnarray}%
where $\Delta _{c}=\Delta _{c}^{0}-g_{cb}q_{s} $ and $\Delta _{m}=\Delta _{m}^{0}+g_{mb}q_{s} $ are the
effective cavity and magnon mode detunings, respectively, which include the
slight shift of frequency. The corresponding linearized QLEs, after ignoring
quantum and thermal noise terms, are:
\begin{figure*}[tbp]
\begin{center}
\includegraphics[width=1.6\columnwidth,height=4in]{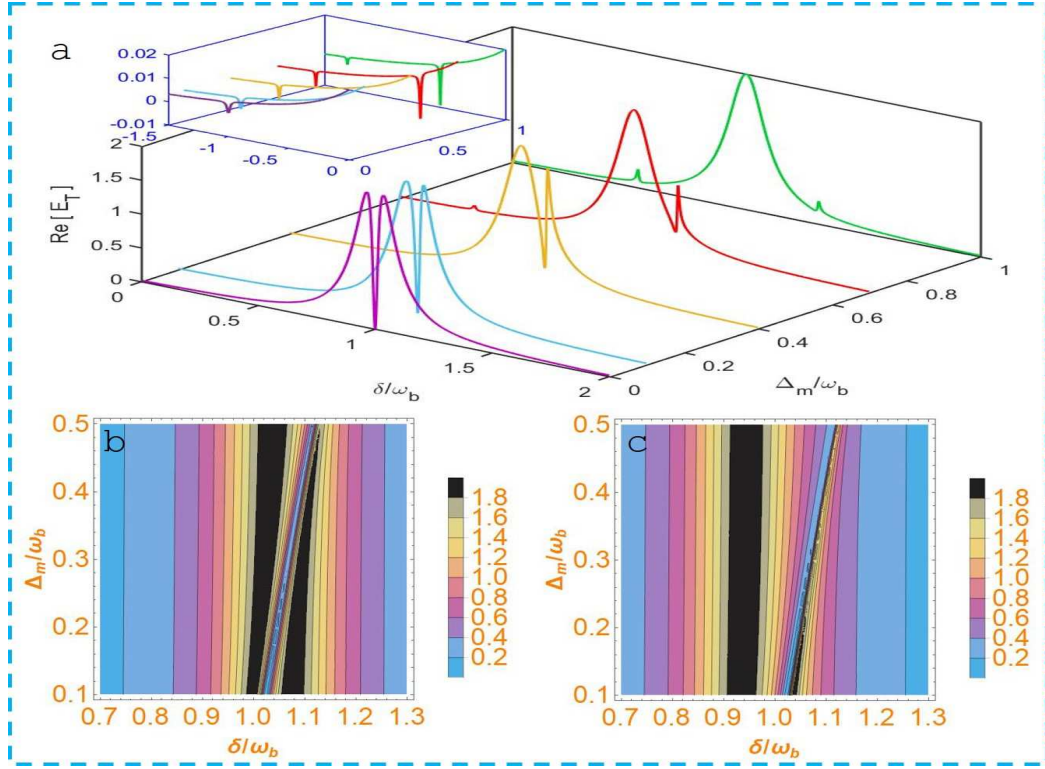}
\end{center}
\caption{(Color Online) (A) The real part Re($\protect\epsilon _{T}$) as a
function of $\protect\delta /\protect\omega _{b}$ and $\Delta _{m}/\protect%
\omega _{b}$ when $\Delta _{c}/\protect\omega _{b}=1$. Contour plot of real
part Re($\protect\epsilon _{T}$) as a function of $\protect\delta /\protect%
\omega _{b}$ and $\Delta _{m}/\protect\omega _{b}$ when (B) $\Delta _{c}=1.05%
\protect\omega _{b}$ and (C) $\Delta _{c}=0.95\protect\omega _{b}$. The rest
of the parameters are same as in Table .1.}
\end{figure*}
\begin{eqnarray}
\delta \dot{q} &=&\omega _{b}\delta p,  \notag \\
\delta \dot{p} &=&\omega _{b}\delta q-\gamma _{b}\delta p-G_{m}\delta
m^{\dagger }-G_{m}^{\ast }\delta m +G_{c}\delta c^{\dagger }+G_{c}^{\ast }\delta c, \notag \\
\delta \dot{c} &=&-(i\Delta _{c}+\kappa _{c})\delta c+iG_{c}\delta
q+\varepsilon _{p}e^{-i\delta t}, \notag \\
\delta \dot{m} &=&-(i\Delta _{m}+\kappa _{m})\delta m-iG_{m}\delta q.
\label{MM}
\end{eqnarray}%
Here, $G_{mb}=\sqrt{2}g_{mb}m_{s} $ and $G_{cb}=\sqrt{2}g_{cb}c_{s} $ are the effective magnomechanical and
optomechanical coupling, respectively. The obtained fluctuation operators
can then be solved by setting: $\delta z=z_{-}e^{-i\delta t}+z_{+}e^{i\delta
t}$ where\ $z_{-}$ and $z_{+}$ (with $z=q,p,c,m$) are much smaller than $%
z_{s}$. It is noteworthy to mention that the fluctuation operators contain
many Fourier components after expansion. Moreover, the high-order terms can
safely be neglected in the limit of weak signal field. Substituting the
above ansatz into Eq. (\ref{MM}) and taking the lowest order in $%
\varepsilon _{p}$, we establish the final solution for the coefficients of
optical response and FWM process as
\begin{equation}
c_{-}=\frac{[\omega _{b}^{2}-\delta ^{2}-i\Theta _{n}]\varepsilon _{p}}{%
[\omega _{b}^{2}-\delta ^{2}-i\Theta _{n}]\Omega _{2}^{c}-i\omega
_{b}G_{cb}^{2}},
\end{equation}%
and
\begin{equation}
c_{+}=\frac{iG_{cb}^{2}\omega _{b}\Theta _{p}\varepsilon _{p}}{[\omega
_{b}^{2}-\delta ^{2}+i\Theta _{p}]\left( \Omega _{1}^{c}\right) ^{\ast
}-i\omega _{b}G_{cb}^{2}},
\end{equation}%
where%
\begin{eqnarray*}
\Theta _{n} &=&\delta \gamma _{b}-\omega _{b}\chi _{12},\ \ \ \ \ \ \ \ \
\Theta _{p}=\delta \gamma _{b}+\omega _{b}\alpha _{12}, \\
\chi _{12} &=&G_{mb}^{2}\chi _{1}+G_{cb}^{2}\chi _{2},\ \ \ \chi
_{1}=[\Omega _{2}^{m}-\Omega _{1}^{m}][\Omega _{1}^{m}\Omega _{2}^{m}]^{-1},
\\
\chi _{2} &=&[\Omega _{1}^{c}]^{-1},\ \ \ \ \ \ \ \ \ \ \ \ \ \ \ \ \ \ \
\alpha _{2}=[\left( \Omega _{2}^{c}\right) ^{\ast }]^{-1}, \\
\alpha _{12} &=&G_{mb}^{2}\alpha _{1}+G_{ab}^{2}\alpha _{2}, \\
\alpha _{1} &=&[\left( \Omega _{1}^{m}\right) ^{\ast }-\left( \Omega
_{2}^{m}\right) ^{\ast }][\left( \Omega _{1}^{m}\right) ^{\ast }\left(
\Omega _{2}^{m}\right) ^{\ast }]^{-1}, \\
\Omega _{1}^{m} &=&\kappa _{m}-i(\Delta _{m}+\delta ),\ \ \ \ \ \ \ \ \
\Omega _{2}^{m}=\kappa _{m}+i(\Delta _{m}-\delta ), \\
\Omega _{1}^{c} &=&\kappa _{c}-i(\Delta _{c}+\delta ),\ \ \ \ \ \ \ \ \ \ \
\ \ \Omega _{2}^{c}=\kappa _{c}+i(\Delta _{c}-\delta ).
\end{eqnarray*}

\section{Output field of OMMS}
To investigate the optical properties of the output field and the FWM
process in OMMS, we start with the following standard input-output relation
\cite{DFW}
\begin{equation}
c_{out}(t)={c}_{in}(t)-\sqrt{2\kappa _{c}}c(t).
\end{equation}%
In the above formula, ${c}_{in}$ and $c_{out}$ are, respectively the input
field and output field operators. Furthermore, the expectation value of the
output field is given by%
\begin{eqnarray}
c_{out}(t) &=&\left( \sqrt{2\kappa _{c}}c_{0}-\frac{\varepsilon _{L}}{\sqrt{%
2\kappa _{c}}}\right) e^{-i\omega _{L}t}  \notag \\
&&+ \left( \sqrt{2\kappa _{c}}c_{-}-\frac{\varepsilon _{p}}{\sqrt{2\kappa
_{c}}}\right) e^{-i(\delta+\omega _{L} )t}  \notag \\
&&+\sqrt{2\kappa _{c}}c_{+}e^{i(\delta-\omega _{L} )t},  \notag \\
&=&\left( \sqrt{2\kappa _{c}}c_{0}-\frac{\varepsilon _{L}}{\sqrt{2\kappa _{c}%
}}\right) e^{-i\omega _{L}t}  \notag \\
&&+\left( \sqrt{2\kappa _{c}}c_{-}-\frac{\varepsilon _{p}}{\sqrt{2\kappa _{c}%
}}\right) e^{-i\omega _{p}t}  \notag \\
&&+\sqrt{2\kappa _{c}}c_{+}e^{i(\omega _{p}-2\omega _{L})t} ,  \label{c1}
\end{eqnarray}%
where the second (third) term corresponds to the components at the
frequencies $\omega _{p}$ ($2\omega _{L}-\omega _{p}$) representing the
optical (FWM) response of the system induced by the anti-Stokes (Stokes)
scattering process. In addition, $c_{out}(t)$ can also be expanded as

\begin{figure}[b!]
\begin{center}
\includegraphics[width=1\columnwidth,height=3.8in]{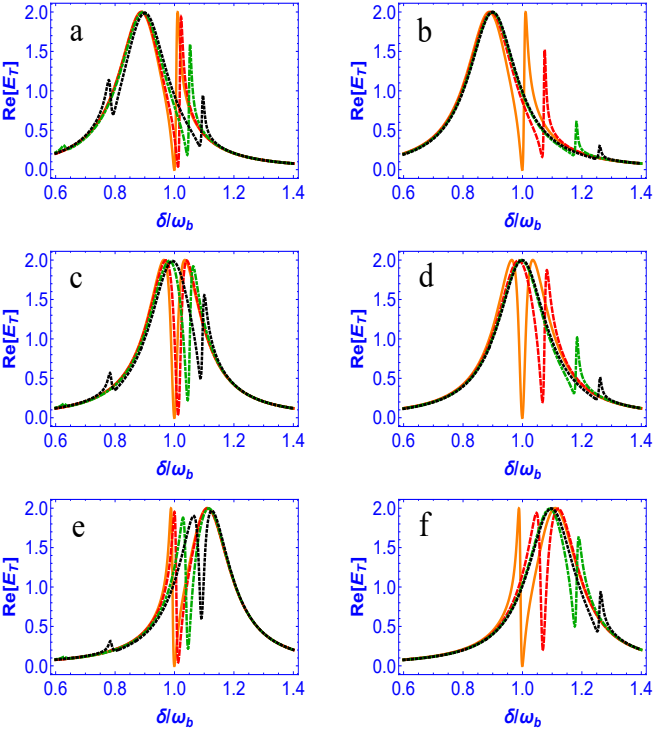}
\end{center}
\caption{(Color Online) The real part Re($\protect\varepsilon _{T}$) as a
function of normalized detuning for different value of $\Delta _{m}$ and $%
G_{mb}$. In figure(a-f), solid orange represents $\Delta _{m}=0$, dashed red
represents $\Delta _{m}=0.3\protect\omega _{b}$, dot-dashed green represents
$\Delta _{m}=0.7\protect\omega _{b}$ and solid gray represents $\Delta
_{m}=0.9\protect\omega _{b}$. We take $G_{mb}=0.2\protect\omega _{b}$ for
the left panel and $G_{mb}=0.5\protect\omega _{b}$ for the right panel.
Moreover, we take $\Delta _{c}=0.9\protect\omega _{b}$ in (a-b), $\Delta
_{c}=\protect\omega _{b}$ in (c-d), $\Delta _{c}=1.1\protect\omega _{b}$ in
(e-f). The rest of the parameters are same as in Table .1.}
\end{figure}
\begin{figure*}[tbp]
\begin{center}
\includegraphics[width=2\columnwidth,height=3in]{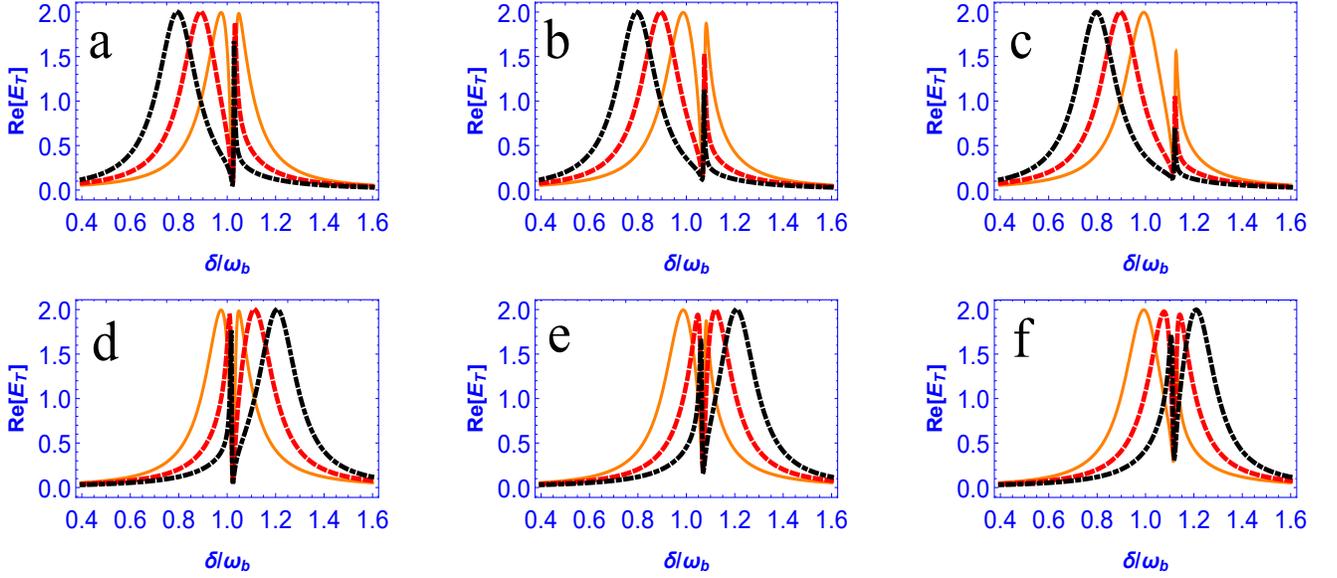}
\end{center}
\caption{(Color Online) The Real part Re($\protect\varepsilon_{T}$) as a
function of normalized detuning for different value of $\Delta_{c}$. In
figure(a-c), solid orange represents $\Delta_{c}=\protect\omega_{b}$, dashed
red represents $\Delta_{c}=0.9\protect\omega_{b}$ and dot-dashed black
represents $\Delta_{c}=0.8\protect\omega_{b}$.In figure(d-f),solid orange
represents $\Delta_{c}=\protect\omega_{b}$, dashed red represents $%
\Delta_{c}=1.1\protect\omega_{b}$ and dot-dashed black represents $%
\Delta_{c}=1.2\protect\omega_{b}$. We take, $\Delta_{m}=0.1\protect\omega%
_{b} $ in (a)(d), $\Delta_{m}=0.3\protect\omega_{b}$ in (b)(e) and $%
\Delta_{m}=0.5\protect\omega_{b}$ in (c)(f). The rest of the parameters are
same as in Table .1.}
\end{figure*}

\begin{eqnarray}
\left\langle c_{out}(t)\right\rangle &=&c_{out,0}e^{-i\omega
_{L}t}+c_{out,-}e^{-i(\delta+\omega _{L} )t}  \notag \\
&&+c_{out,+}e^{i(\delta-\omega _{L} )t},  \label{c2}
\end{eqnarray}%
Comparing Eq. (\ref{c1}) and Eq. (\ref{c2}), we easily arrive at
\begin{eqnarray}
c_{out,-} &=&\sqrt{2\kappa _{c}}c_{-}-\frac{\varepsilon _{p}}{\sqrt{2\kappa
_{c}}},  \notag \\
c_{out,+} &=&\sqrt{2\kappa _{c}}c_{+}.
\end{eqnarray}%
Finally, we obtain the following total output field relation at the probing
frequency $\omega _{p}$
\begin{eqnarray}
\varepsilon _{T} &=&\frac{\sqrt{2\kappa _{c}}c_{out,-}+\varepsilon _{p}}{%
\varepsilon _{p}}=\frac{2\kappa _{c}c_{-}}{\varepsilon _{p}}=\Lambda+i\tilde{\Lambda},
\end{eqnarray}%
where $\Lambda$=Re[$\varepsilon _{T}$] and $\tilde{\Lambda}$=Im[$\varepsilon
_{T}$] are, respectively, represent the in-phase and out-of-phase
quadratures of the output probe field. In addition, one can obtained the
absorptive (dispersive) behavior of the output probe field from $\Lambda$($%
\tilde{\Lambda}$) \cite{AMEPJD}.

Similarly, at Stokes scattering, the FWM intensity can be obtained from
output stokes field as
\begin{equation}
\varepsilon _{FWM}=\left\vert \frac{\sqrt{2\kappa _{c}}c_{out,+}}{%
\varepsilon _{p}}\right\vert ^{2}=\left\vert \frac{2\kappa _{c}c_{+}}{%
\varepsilon _{p}}\right\vert ^{2}.
\end{equation}
\begin{table}
\centering
\begin{tabular*}{0.5\textwidth}{@{\extracolsep{\fill}}|l c c |}
\hline
Parameters & Symbol & Value  \tabularnewline
\hline
Decay rate of the cavity mode  & $\kappa_{c}$&$0.1\omega_{b}$ \tabularnewline

Decay rate of the magnon mode  & $\kappa_{m}$&$\simeq0.1\kappa_{c}$ \tabularnewline

Mechanical damping rate & $\gamma_{b}$ & $ 10^{-5}\omega_{b}$ \tabularnewline

Effective Cavity coupling  & $G_{cb}$ & $0.05\omega_{b}$ \tabularnewline

Effective magnon coupling  & $G_{mb}$ & $0.5\omega_{b}$ \tabularnewline
\hline
\end{tabular*}
\caption{The parameters used in our calculations, taken from recent the experiments in Ref \cite{Tang,disp2,AMPS}.} 
\label{table:pvalue}
\end{table}
\section{Numerical results and discussions}
\subsection{Fano-type optical response}
In this section, we present the frequency response and the FWM process in
OMMS by employing the system parameters. For the present OMMS, we have
selected parameters which are well experimentally realizable and are given
in Table 1 \cite{Tang,disp2,AMPS}. The practical feasibility of these
parameters is given in \cite{{AMPS,ASIJTP,AMPS2}}. It is noteworthy that the
cavity mode is weakly driven as compared to the driven magnon, leading to $%
\left\vert c_{s} \right\vert ^2 \ll \left\vert
m_{s} \right\vert^2  $ and therefore, $G_{cb}\ll
G_{mb} $. Hence, one can safely claim that the megnetostrictive interaction
is paramount over optomechanical interaction, i.e., $q_{s}\simeq -g_{cb}\frac{\left\vert m_{s}
\right\vert ^{2}}{\omega _{b}}$ (See Eq. (\ref{SS})). In addition, it has
already been proved that effective magnomechanical coupling, $G_{mb}$ is
almost $10$ times greater than effective magnomechanical coupling, $G_{cb}$
\cite{ASIJTP,ZRCS}. Therefore for simplicity, we have assumed $%
G_{mb}=(0.2-0.5)\omega_{b}$ and $G_{cb}=0.05\omega_{b} $.

Fano resonances emerge due to the interaction between the cavity/magnon mode
and the vibrational mode of the resonator. We have realized the resonances
in the output spectra viz. $\varepsilon _{T}=\Lambda+i\tilde{\Lambda}$.
Since OMIT originates by the constructive and destructive interference of
different frequency contributions, it is well known that we can expect Fano
profiles under some conditions \cite{KQGS}.
We first note that when the magnon (or cavity) detuning is out of resonance
with the vibrational frequency, for instance $\Delta_{m}<\omega_{b}$ and $%
\Delta_{c}\neq\omega_{b}$ (or even $\Delta_{c}=\omega_{b}$), consequently
Fano profiles emerge in OMIT spectra, in agreement with previous reports
\cite{KQGS,MYPOM1}. To discuss the emergence and tunability of Fano profiles
in OMMS, we first show the absorption spectra as a function of normalized
detuning $\delta/\omega_{b}$ for a fixed value of magnomechanical coupling
rate $G_{mb}$, when $\Delta_{c}=\omega_{b}$ in Fig. 2. Starting from $%
\Delta_{m}/\omega_{b}=0$, we tuned the frequency $\Delta_{m}$ in such a way
that it approached the vicinity of the mechanical mode frequency $\omega_{b}$
(see the green curve in Fig. 2 (a)).
In the absence of magnon interaction, two dips in absorption can be found,
which appear around $\delta=\pm\omega_{b}$ as shown by the magenta line in
Fig. Fig. 2 (a) (Also in inset). This is because in the present system, we
have taken extremely low value of the damping rate and therefore the lowest
value of the output field amplitude exist when $\omega _{b}^{2}-\delta
^{2}\simeq0$ (or $\delta=\pm\omega _{b}$). Consequently, we observe a
typical OMIT windows around these points. However, the magnomechanical
interaction involve to exhibit asymmetric Fano profiles as shown by the
yellow, red and green lines in Fig. 2(a). In addition, we obtain a curve
almost similar to Lorentzian when $\Delta_{m}=\omega_{b} $. Hence, the
interaction of magnomechanical interaction leads to enhance the absorption
around $\delta=\omega_{b}$ around $\Delta_{m}=\omega_{b} $. However the peak
of the Fano profiles gradually decrease with the increase of $\Delta_{m}$.
The inset 3D figure in Fig. (a) exhibits the same absorption spectra in the
blue-detuned region, which shows the small amplification of each curve
occurs where we observe the dip in the red-detuned region. For example, the
absorption dips (amplification dips) of the green line in red-detuned region
occur at $\delta=0.55\omega_{b}$ ($\delta=-0.55\omega_{b}$) and $%
\delta=1.3\omega_{b}$ ($\delta=-1.3\omega_{b}$) (-ve values can be seen in
inset of Fig. 2(a)). To further shed light on the Fano profiles, we show the
Fano profile spectra for a specific range of magnon detuning as density
plots in Fig. 2 (b) and Fig. 2 (c) at $\Delta_{c}=1.05\omega_{b}$ and $%
\Delta_{c}=0.95\omega_{b}$ respectively. The Fano profiles become more
prominent when we slightly shift the cavity detuning to higher frequency
(see Fig. 2 (b)) as compared to the Fano profile spectra for the lower
cavity detuned region (see Fig. 2 (c)).

As already discussed, the existence of Fano profiles in OMMS is due to the
nonresonant interactions of magnon (or cavity) mode, the symmetry of the
typical OMIT window is transfigured into asymmetric Fano profiles.
Therefore, in the present OMMS, we investigate the Fano profile which can be
tuned by the effective magnomechanical coupling strength $G_{mb}$ and magnon
detuning $\Delta_{m}$ for both resonance ($\Delta_{c}=\omega_{b}$) as well
as a non-resonance case ($\Delta_{c}\neq\omega_{b}$). In Fig. 3, we show a
typical behavior of absorption spectra as a function of the normalized
probe-coupling detuning for different values of $\Delta_{m}$ and $G_{mb}$.
We obtained a typical OMIT window in the absence of magnon interaction as
show by the orange solid line in Fig. 3 (c-d). However, the asymmetry of the
absorption spectra incenses with the increase of $\Delta_{m}$ as shown by
the dashed red, dot-dashed green and solid black lines. Similarly, Fig. 3
(a-b) shows the absorption spectra when $\Delta_{c}=0.9\omega_{b}$ and Fig.
3 (e-f) shows the absorption spectra when $\Delta_{c}=1.1\omega_{b}$. The
height of the Fano profiles decrease with the increase of $\Delta_{m}$. In
addition, one can see that the width between the central absorption peak and
sharp Fano profile for left panel of Fig. 3, is not prominently increase
while for the right panel, it sufficiently increases.
Therefore, it is clear from the above discussion, that $\Delta_{m}$ ($G_{mb}$%
) controls the peak/dip (width) of the Fano profile.

In order to observe the combined effect of magnon and cavity detunings, we
plot the absorption spectra by varying $\Delta_{m}/\omega_{b}$ and $%
\Delta_{c}/\omega_{b}$ in Fig. 4. We now present the absorption profile as a
function of normalized detuning $\delta/\omega_{b}$ at the fixed effective
coupling strengths $G_{mb}$ and $G_{cb}$. For a fixed value of $%
\Delta_{m}/\omega_{b}$, we continuously tuned the cavity detuning from $%
\Delta_{c}=0.8\omega_{b}$ to $\Delta_{c}=1.2\omega_{b}$ as shown in Fig. 4.
It is noted that the Lorentzian type peak move left (right) as cavity
detuning $\Delta_{c}/\omega_{b}$ decreases (increases). However, the peak of
the sharp Fano profile depends mainly on the magnon detuning $%
\Delta_{m}/\omega_{b}$ i.e., peak (dip) decreases (move slightly upward)
with the increase of $\Delta_{m}/\omega_{b}$.

\subsection{Four-wave mixing process}
\begin{figure}[b!]
\begin{center}
\includegraphics[width=1\columnwidth,height=5in]{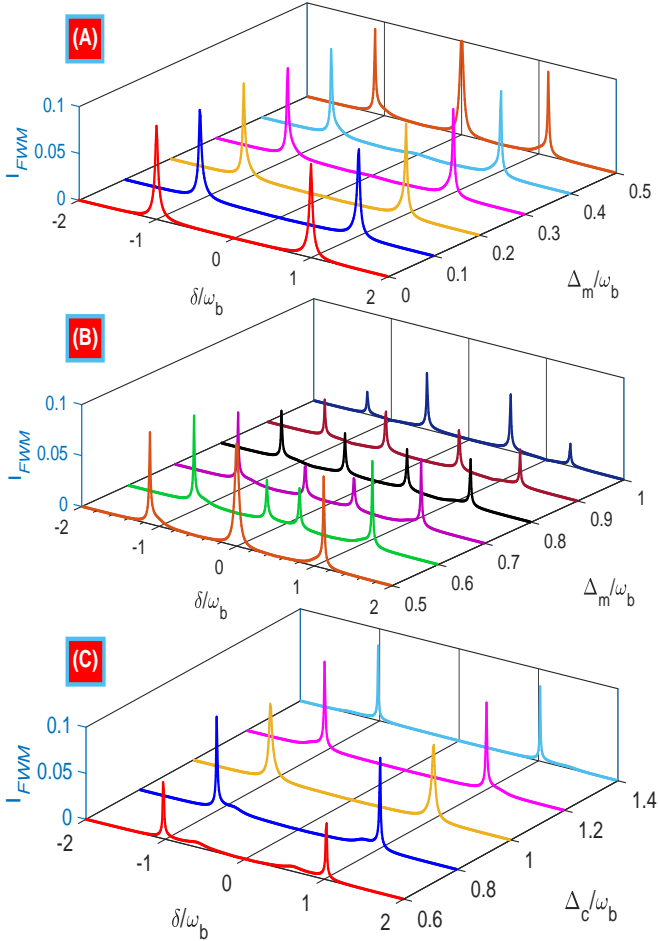}
\end{center}
\caption{(Color Online) FWM intensity $I_{FWM}$ as a function of $\protect%
\delta/\protect\omega_{b}$ and $\Delta_{m}/\protect\omega_{b}$. We take $%
\Delta_{c}=\protect\omega_{b}$ in (a-b) and $\Delta_{m}=0.4\protect\omega%
_{b} $ in (c). The rest of the parameters are same as in Table .1.}
\end{figure}
\begin{figure}[b!]
\begin{center}
\includegraphics[width=1\columnwidth,height=3.5in]{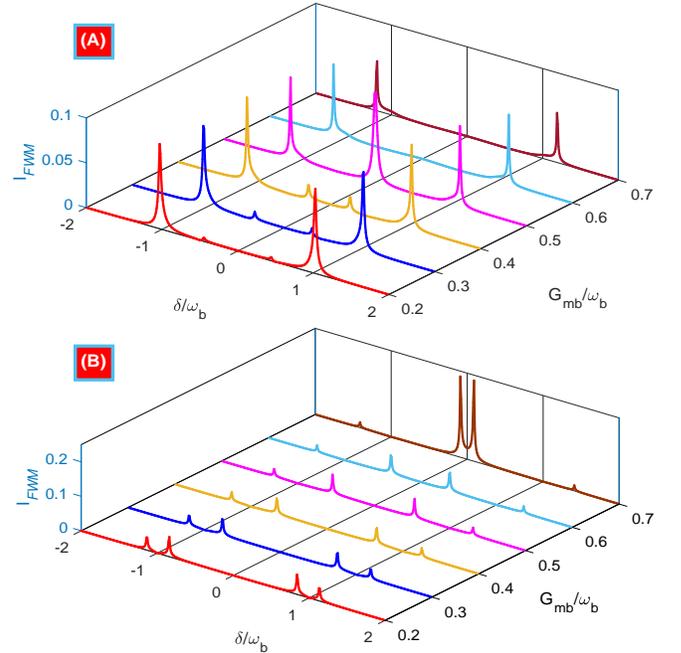}
\end{center}
\caption{(Color Online) FWM intensity $I_{FWM}$ as a function of (A) $%
\Delta_{c}/\protect\omega_{b}$ and $\protect\delta/\protect\omega_{b}$ and
(B) $G_{mb}/\protect\omega_{b}$ and $\protect\delta/\protect\omega_{b}$. We
take $\Delta_{m}=0.5\protect\omega_{b}$ in (A) and $\Delta_{m}=\protect\omega%
_{b}$ in (B). The rest of the parameters are same as in Table .1.}
\end{figure}
Compared with the typical cavity optomechanical system comprising a
mechanical resonator coupled with an optical cavity via radiation pressure
force, the OMMS under consideration allows a greater controllability on the
FWM process. Here, we concentrate on different parameters to tune the FWM
signals, i.e., the cavity/magnon detuning, the decay rate and effective
magnomechanical coupling strength. Fig. 5(A) plots the FWM spectrum as a
function of the pump-probe detuning for different values of the magnon
detuning. It is shown in Fig. 5(A) that the two peaks of FWM signals are
obtained exactly at $\delta=\pm\omega_{b}$, where OMMS exhibits
transparency. The peaks at $\delta=\pm\omega_{b}$ represent the intensity of
the stokes field produced by the coupling field with the vibrational mode
frequency. In addition, for $\Delta_{m}<0.4\omega_{b}$, the FWM signal
remains almost the same as shown in Fig. 5(A). However, one can see an extra
peak located at $\delta=0$ along with the two peaks located at $\delta\simeq
\pm0.1\omega_{b}$ appear when $\Delta_{m}=0.5\omega_{b}$. Therefore, one can
infer that the middle peak arises owing to the strong magnomechanical
interaction. In addition, for $\Delta_{m}\geq0.6\omega_{b}$, the peak at $%
\delta=0$ further split to create two peaks at $\delta=\pm0.2\omega_{b}$
along with two peaks at $\delta=\pm1.2\omega_{b}$. However, the strength of
the two peaks at $\delta=\pm0.2\omega_{b}$ sufficiently smaller than the two
peaks around $\delta=\pm\omega_{b}$. It can be seen that as the
magnomechanical interaction becomes strong (i.e., for $\Delta_{m}\geq0.6%
\omega_{b}$), the two peaks at $\delta=\pm0.2\omega_{b}$ become heighten as
compared to the two peaks around $\delta=\pm\omega_{b}$ as shown in Fig.
5(B). This effect becomes more prominent in Fig. 6(a-b).
Furthermore, the width between the peaks gets broader by increasing $%
\Delta_{m}$ as shown in Fig. 5(B). It is important to mention here that for
increasing value of $\Delta_{m}$, the FWM signals at $\delta=\omega_{b}$ ($%
\delta=-\omega_{b}$) start moving to the right(left). Therefore, we infer
that the signal at $\delta=\pm\omega_{b}$ is completely suppressed at due to
the destructive interference (see the dark blue line). Fig. 5(C) plots the
FWM spectrum as a function of the pump-probe detuning for different value of
the cavity detuning which clearly shows the FWM signal becomes strong as we
increase $\Delta_{c}$ and the width between the peaks remains constant.

Figure 6(A) shows the variation of the FWM spectrum as a function of the
pump-probe detuning for different values of the effective magnon coupling
strength $G_{mb}$ when $\Delta_{m}=0.5\omega_{b}$. It can be shown that the
two less intensity middle peaks gets stronger and closer until $%
G_{mb}\leq0.4\omega_{b}$. As the magnomechanical interaction becomes
stronger, the two middle less intensified peaks then merged to form a single
intensified peak for $G_{mb}=0.5\omega_{b}$. Further enhancement of $G_{mb}$
not only brings the peak at $\delta=0$ to an end but also decrease the
intensity of two side peaks. Figure 6(B) shows the variation of the FWM
spectrum as a function of the pump-probe detuning for different value of the
effective magnon coupling strength $G_{mb}$ when $\Delta_{m}=\omega_{b}$.
One can observe that the two distant peaks move further apart while the two
middle peaks come closer by increasing the value of the $G_{mb}$. In
addition, the strength of the two distant peaks (middle peaks) decrease
(increase) with the enhancement of $G_{mb}$. From the above discussion, one
can easily conclude that the FWM signals can be tuned (enhanced) by varying
different system's parameters.

The FWM process has been shown in cavity optomechanics, which strongly
depends on the mean photon number. Varying the cavity decay rate $\kappa_{c}$
(magnon decay rate $\kappa_{m}$) will directly modify the mean photon number
(magnon number). Therefore, we plot the FWM signal by varying the cavity and
magnon decay rates. As shown in Fig. 7(A), the FWM signal is strong and the
two peaks can be seen at $\delta=\pm\omega_{b}$. When the cavity decay rate
is decreased, the two peaks of the FWM signal is sufficiently reduced. In
addition, the two peaks located at $\delta=\pm\omega_{b}$ transfigured to
four diminished peaks located at $\delta=\pm0.95\omega_{b}$ and $%
\delta=\pm1.05\omega_{b}$. Similarly, in Fig. 7(B), we investigate the
influence of the magnon decay rate $\kappa_{m}$ on the FWM intensity. Figure
7(B) shows that the FWM intensity is enhanced sufficiently with the increase
of the magnon decay rate $\kappa_{m}$. Hence, one can infer that cavity and
magnon decay rates have a substantial impact on the FWM process \cite{FWM7}.
\subsection{Experimental Feasibility}
In the following, we discuss the experimental feasibility of our proposed scheme. There exist two subsystems: (i) Magnomechanical part (ii) Optomechanical part. The adopted YIG micro-bridge (the magnomechanical subsystem) was based on the experimental demonstration \cite{Heyroth}. However, the experimental demonstration of the optomechanical part has been demonstrated in many research papers \cite{OM1,OM2}. As for as experimental parameters are concerned, we have chosen parameters which are can easily be achieved using present-day technology. In addition, we have considered the parametric regime which is
valid only when the magnon occupation number is significantly smaller than the total number of spin i.e., $m^{\dag}m << 2N_{s}=5N$, where 5N is the total spin number of the $Fe^{3+}$ ion in YIG $s=5/2$. Our numerical results show that $N$ is of the order of $3.5\times10^{16}$ and $\Omega=712 THz$ \cite{Tang,AMPS}. It has been observed in the present study that these
conditions are fulfilled and our scheme is feasible for strong magnon-phonon coupling. This is due to the strong magnon pump field, which causes unwanted nonlinear effects. However, as shown in recent experiments, a larger value of magnon-phonon coupling would result in a working regime, where we can drop the higher order terms due to a strongly driven magnomechanical system \cite{AMPS}. From the above discussion, we infer that we have chosen an experimentally feasible working regime. In near future,
any experiment employing a planar cross-shaped cavity coupled to a magnon or can allow the implementation of the proposed scheme using a co-planar waveguide \cite{{RJW,LGQ}}.
\begin{figure}[tbp]
\begin{center}
\includegraphics[width=1\columnwidth,height=3.5in]{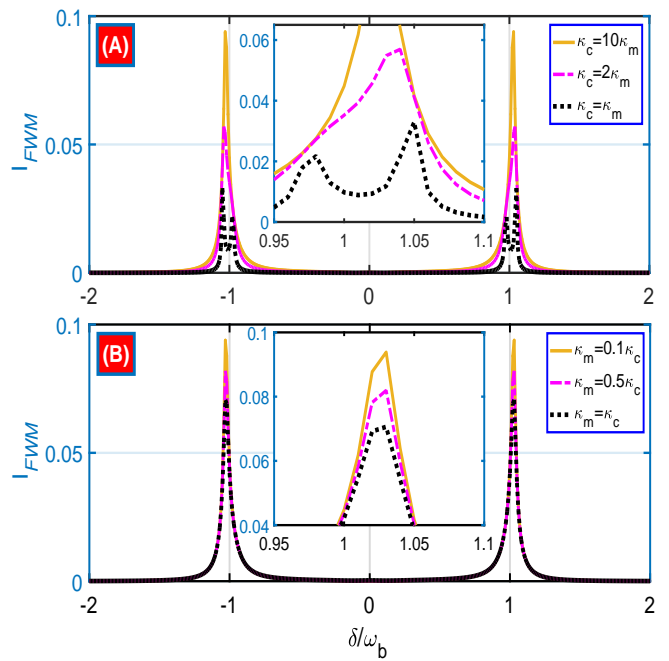}
\end{center}
\caption{(Color Online) FWM intensity $I_{FWM}$ as a function of $\protect%
\delta/\protect\omega_{b}$ for different values of (A) $\protect\kappa_{c}$
and (B) $\protect\kappa_{m}$. We take $\protect\kappa_{m}=0.01\protect\omega%
_{b}$ in (A) and $\protect\kappa_{c}=0.1\protect\omega_{b}$ in (B). The rest
of the parameters are same as in Table .1.}
\end{figure}
\section{Conclusion}
In conclusion, we have introduced a model consisting of an optomechanical system directly coupled with a ferromagnetic material. The deformation of the mechanical mode is taken simultaneously due to an  optomechanical coupling and the magnetoelasticity of the ferromagnet. We report that the magnetostrictively induced displacement evolves to transfigure the standard OMIT window to Fano profiles in the output field at the probe frequency. The asymmetry of the Fano profiles can be well tuned/controlled
by appropriate adjusting the magnon (cavity) detuning and the effective magnomechanical coupling. Remarkably, it is found that the magnetoelastic interaction also gives rise to the FWM phenomenon. FWM intensity can be resonantly tuned by the system's parameters. For example, modulating the effective magnomechanical coupling can realize conversion among a double and triple peak signals. Similarly, in the case of magnon detuning, FWM exhibits the characteristics of symmetric double, triple and even quadruple peak signals. In addition, one can also control the strength of the FWM signals by employing the magnon (cavity) detuning and effective magnomechanical coupling. Moreover, the FWM spectrum exhibits suppressive behavior upon increasing (decreasing) the magnon (cavity) decay. The present OMMS opens new perspectives in highly sensitive detection and quantum information processing, which shall be addressed elsewhere.
\section*{Appendix A. Magnomechanical coupling}
The magnetoelastic coupling is fully explained by the interaction between the elastic strain and magnetization of the magnetic material \cite{RBWD,Kittel}. The magnetoelastic energy density for a cubic crystal is
given by \cite{KIT}
\begin{eqnarray}
F_{me} &=&\frac{2B_{1}}{M_{s}^{2}}\left( M_{x}M_{y}\epsilon
_{xy}+M_{x}M_{z}\epsilon _{xz}+M_{y}M_{z}\epsilon _{yz}\right) \notag \\
&& +\frac{B_{2}}{M_{s}^{2}}\left( M_{x}^{2}\epsilon
_{xx}+M_{y}^{2}\epsilon _{yy}+M_{z}^{2}\epsilon _{zz}\right),  \label{A1}
\end{eqnarray}%
where the magnetoelastic coupling coefficients are represented by $B_{1}$
and $B_{2}$. $M_{s}$ being the saturation magnetization. In addition,
magnetization components are represented by $M_{x}$, $M_{y}$ and $M_{z}$.
Furthermore, $\epsilon _{nm}$ ($n,m\in {x,y,z}$) defines the strain tensor
of the cubic magnetic crystal.

Now, the magnetization can be quantized using the Holstein-Primakoff transformation
\begin{equation}
m=\sqrt{\frac{V}{2\hbar \gamma M_{s}}}\left( M_{x}-iM_{y}\right) ,
\end{equation}%
where $V$ is the volume of the cubic crystal and $m$ defines the magnon mode
operator. After rearranging, we obtain
\begin{eqnarray}
M_{x} &=&\sqrt{\frac{\hbar \gamma M_{s}}{2V}}\left( m+m^{\dagger }\right) ,
\notag \\
M_{y} &=&i\sqrt{\frac{\hbar \gamma M_{s}}{2V}}\left( m-m^{\dagger }\right) ,
\end{eqnarray}%
and%
\begin{equation}
M_{z}=\left[ M_{s}^{2}-M_{x}^{2}-M_{y}^{2}\right] ^{\frac{1}{2}}\simeq M_{s}-%
\frac{\hbar \gamma }{V}m^{\dagger }m.
\end{equation}%
Putting the value of $M_{x,y,z}$ in Eq. (\ref{A1}) and then integrating over
the volume of the cubic magnetic crystal, the first term in Eq. (\ref{A1})
yields the semiclassical magnetoelastic Hamiltonian,
\begin{eqnarray}
H_{a} &\simeq &\frac{B_{1}}{M_{s}}\frac{\hbar \gamma }{V}m^{\dagger }m\int
dl^{3}\left( \epsilon _{xx}+\epsilon _{yy}-2\epsilon _{zz}\right)  \notag \\
&&+\frac{B_{1}}{M_{s}}\frac{\hbar \gamma }{2V}\left( m^{2}+m^{\dagger
2}\right) \int dl^{3}\left( \epsilon _{xx}-\epsilon _{yy}\right)  \notag \\
&&+\frac{B_{1}}{M_{s}^{2}}\frac{\hbar ^{2}\gamma ^{2}}{V^{2}}m^{\dagger
}mm^{\dagger }m\int dl^{3}\epsilon _{zz},  \label{Ha}
\end{eqnarray}%
and the second term of magnetoelastic energy density from Eq. (\ref{A1})
yields the Hamiltonian
\begin{eqnarray}
H_{b} &\simeq &\frac{B_{2}}{M_{s}}\frac{\hbar \gamma }{V}\left(
m^{2}-m^{\dagger 2}\right) \int dl^{3}\epsilon _{xy}  \notag \\
&&+\frac{2B_{2}}{M_{s}}\sqrt{\frac{\hbar \gamma M_{s}}{2V}}\left[ M_{s}-%
\frac{\hbar \gamma }{V}m^{\dagger }m\right]  \notag \\
&&\times \left[ m\int dl^{3}\left( \epsilon _{xz}+i\epsilon _{yz}\right) +H.c%
\right] .  \label{Hb}
\end{eqnarray}%
Hence, the total magnetoelastic Hamiltonian, which is sum of  $H_{a}$ and $H_{b}$, implies
total magnon-phonon interactions

The magnetoelastic displacement can be defined as:
\begin{equation}
\vec{u}\left( x,y,z\right) =\sum\limits_{l,m,n}d^{(l,m,n)}\vec{\chi}%
^{(l,m,n)}\left( x,y,z\right) ,  \label{U}
\end{equation}%
where $\vec{\chi}^{(l,m,n)}\left( x,y,z\right) $ ($d^{(l,m,n)}$) denotes the
normalized displacement eigenmode (the corresponding amplitude).
\begin{equation}
d^{(l,m,n)}=d_{zpm}^{(l,m,n)}\left( b_{l,m,n}+b_{l,m,n}^{\dagger }\right) ,
\label{D}
\end{equation}%
where $d_{zpm}^{(l,m,n)}$ represents the amplitude of the zero-point motion.
In addition, $b_{l,m,n}$ and $b_{l,m,n}^{\dagger }$ are, respectively the
corresponding phononic annihilation and creation operators. Substituting
Eqs. (\ref{U}) and (\ref{D}) in Eq. (\ref{Ha}), the dispersive kind of
interaction can be obtained,
\begin{equation}
H_{a}=\hbar \sum\limits_{l,m,n}g_{mb}^{(l,m,n)}m^{\dagger }m\left(
b_{l,m,n}+b_{l,m,n}^{\dagger }\right) ,
\end{equation}%
where $g_{mb}^{(l,m,n)}$ represents the dispersive coupling strength between
magnon and phonon, is given by
\begin{eqnarray}
g_{mb}^{(l,m,n)} &=&\frac{B_{1}}{M_{s}}\frac{\gamma }{V}\int
dl^{3}d_{zmp}^{(l,m,n)}  \notag \\
&&\times \left[ \frac{\partial \chi _{x}^{(l,m,n)}}{\partial x}+\frac{%
\partial \chi _{y}^{(l,m,n)}}{\partial y}-2\frac{\partial \chi _{z}^{(l,m,n)}%
}{\partial z}\right] .
\end{eqnarray}%
In case of a one dimensional mechanical mode oscillation, we obtain the
simplified hamiltonian as:
\begin{equation}
H_{a}=\hbar g_{mb}m^{\dagger }m\left( b+b^{\dagger }\right) ,
\end{equation}%
which is the same as the first part of $H_{i}$ (see Eq. (\ref{Hi})), since $%
q=\frac{1}{\sqrt{2}}\left( b+b^{\dagger }\right) $. Note that we use the
condition of low-lying magnon excitations to derive the above and therefore,
the magnon excitation $m^{\dagger }m$ second order term (third line of Eq. (%
\ref{Ha}) ) can be safely eliminated.

\section*{Data availability}

All data used during this study are available within the article.


\begin{thebibliography}{99}
\bibitem{Mey} M. Aspelmeyer, P. Meystre, and K. Schwab, ``Quantum
optomechanics". Phys. Today \textbf{65}(7) 29 (2012).

\bibitem{Fio} C. H. Dong, V. Fiore, M.C. Kuzyk, and H. L. Wang,
``Optomechanical dark mode". Science.; \textbf{338} 1609 (2012).

\bibitem{Jalil} J. K. Moqadam, R. Portugal, M. C. de Oliveira, ``Quantum
walks on a circle with optomechanical systems". Quantum Inf Process \textbf{%
14}, 3595 (2015).

\bibitem{Teu} J. D. Teufel, T. Donner, M. A. Castellanos-Beltran, J. W.
Harlow and K. W. Lehnert, ``Nanomechanical motion measured with an
imprecision below that at the standard quantum limit". Nature nanotechnology
; \textbf{4}(12): 820 (2009).

\bibitem{AliM} A. Motazedifard, A. Dalafi and M. H. Naderi, ``Ultraprecision
quantum sensing and measurement based on nonlinear hybrid optomechanical
systems containing ultracold atoms or atomic Bose%
\"{}%
CEinstein condensate" AVS Quantum Sci. \textbf{3}, 024701 (2021)

\bibitem{Qian} Y. Jiao, H. Lu, J. Qian, Y. Li and H. Jing, ``Nonlinear
optomechanics with gain and loss: amplifying higher-order sideband and group
delay". New J. Phys. ; \textbf{18}(8):083034 (2016).

\bibitem{Lud} L. M. Ludwig, A. H. Safavi-Naeini, O. Painter and F.
Marquardt, ``Enhanced quantum nonlinearities in a two-mode optomechanical
system". Phys. Rev. Lett. ; \textbf{109}(6):063601 (2012).

\bibitem{SS1} S. K. Singh and C. H. Raymond Ooi, \textexclamdown
${{}^\circ}$%
Quantum correlations of quadratic optomechanical oscillator, J. Opt. Soc.
Am. B \textbf{31}, 2390 (2014).

\bibitem{SS2} S. K. Singh and S. V. Muniandy, \textexclamdown
${{}^\circ}$%
Temporal Dynamics and Nonclassical Photon Statistics of Quadratically
Coupled Optomechanical Systems,\textexclamdown $\pm$ Int. J. Theor. Phys.
\textbf{55}, 287 (2016).

\bibitem{AM0} A. Sohail, M. Rana, S. Ikram, T. Munir, T. Hussain, R. Ahmed,
C. S. Yu. ``Enhancement of mechanical entanglement in hybrid optomechanical
system". Quantum Information Processing ; \textbf{19}(10):372 (2020).

\bibitem{MYP} A. Sohail, \textit{et al.}, ``Enhanced entanglement induced by
Coulomb interaction in coupled optomechanical systems". Phys. Scr. \textbf{95%
}:035108 (2020).

\bibitem{SKS} S. K Singh , J. X. Peng, M. Asjad and M. Mazaheri,
``Entanglement and coherence in a hybrid Laguerre-Gaussian rotating cavity
optomechanical system with two-level atoms". J. Phys. B: At. Mol. Opt. Phys.
\textbf{54}:215502 (2021).

\bibitem{BTT} Berihu Teklu, Tim Byrnes, Faisal Shah Khan. ``Cavity-induced
mirror-mirror entanglement in a single-atom Raman laser". Phys. Rev. A
\textbf{97}:023829 (2018).

\bibitem{Arva} A. Arvanitaki and A. A. Geraci, ``Detecting high-frequency
gravitational waves with optically levitated sensors". Phys. Rev. Lett.
\textbf{110}(7):071105 (2013).

\bibitem{AliM2} M. S. Ebrahimi, A. Motazedifard, and M. Bagheri Harouni,
``Single-quadrature quantum magnetometry in cavity electromagnonics"
Physical Review A \textbf{103} (6), 062605 (2021)

\bibitem{OM1} S. Weis, et al. Optomechanically induced transparency. Science
\textbf{330}, 1520 (2010).

\bibitem{OM2} A. H. Safavi-Naeini, et al. Electromagnetically induced
transparency and slow light with optomechanics. Nature (London) \textbf{472}%
, 69 (2011).

\bibitem{MYPOM1} A. Sohail, Y. Zhang, G, Bary, C. S. Yu. ``Tunable
Optomechanically Induced Transparency and Fano Resonance in Optomechanical
System with Levitated Nanosphere". IJTP \textbf{57}(9) 2814 (2018).

\bibitem{MYPOM2} A. Sohail, R. Ahmed, C. S. Yu. ``Switchable and Enhanced
Absorption via Qubit-Mechanical Nonlinear Interaction in a Hybrid
Optomechanical System". IJTP \textbf{60} 739 (2021).


\bibitem{Faraz} F. Monifi1, J. Zhang, B. Peng1, Yu. Liu, F. Bo, Franco Nori
and L. Yang, ``Optomechanically induced stochastic resonance and chaos
transfer between optical fields". Nature Photon. \textbf{10}(6):399 (2016).

\bibitem{Imam} M. Fleischhauer, A. Imamoglu, and J. P. Marangos,
``Electromagnetically induced transparency: Optics in coherent media".
Reviews of modern physics \textbf{77}(2):633 (2005).

\bibitem{Saif} S. K. Singh, M. Parvez, T. Abbas, J. X. Peng, M. Mazaheri, M.
Asjad, ``Tunable optical response and fast (slow) light in optomechanical
system with phonon pump". Phys. Lett. A \textbf{442} 128181 (2022).

\bibitem{Saiff} S. K. Singh, M. Asjad and C. H. Raymond Ooi, ``Tunable
optical response in a hybrid quadratic optomechanical system coupled with
single semiconductor quantum well". Quantum Inf. Process. 21 47 (2022).

\bibitem{Yuri} A. E. Miroshnichenko, S. Flach, and Y. S. Kivshar, ``Fano
resonances in nanoscale structures". Rev. Mod. Phys. \textbf{82}(3):2257
(2010).

\bibitem{Fano} U.Fano, ``Effects of configuration interaction on intensities
and phase shifts". Phys. Rev. \textbf{124}(6):1866 (1961).

\bibitem{Gav} M. O. Scully, S. Y. Zhu, and A. Gavrielides, ``Degenerate
quantum-beat laser: Lasing without inversion and inversion without lasing".
Phys. Rev. Lett. \textbf{62}(24):2813 (1989).

\bibitem{Kuh} A. B\"{a}rnthaler, S. Rotter, F. Libisch, J. Burgdorfer, S.
Gehler, U. Kuhl, and H. J. Stockmann, St\"{o}ckmann H J. ``Probing
decoherence through Fano resonances". Phys. Rev. Lett. \textbf{105}%
(5):056801 (2010).

\bibitem{Chap} M. O. Scully, K. R. Chapin, K. E. Dorfman, M. B. Kim, and A.
A. Svidzinsky, ``Quantum heat engine power can be increased by noise-induced
coherence". Proc. Natl. Acad. Sci. \textbf{108}(37): 15097 (2011).

\bibitem{Tass} P. Tassin, L. Zhang, R. Zhao, A. Jain, T. Koschny, and C. M.
Soukoulis, ``Electromagnetically induced transparency and absorption in
metamaterials: the radiating two-oscillator model and its experimental
confirmation". Phys. Rev. Lett. \textbf{109}(18):187401 (2012).

\bibitem{Alim2} Ali Motazedifard, A. Dalafi, M. H. Naderi, ``Negative cavity
photon spectral function in an optomechanical system with two
parametrically-driven mechanical oscillators" arXiv:2205.15314 (2022).

\bibitem{Gal} B. Gallinet and O. J. F. Martin, ``Ab initio theory of Fano
resonances in plasmonic nanostructures and metamaterials". Phys. Rev. B
\textbf{83}(23):235427 (2011).

\bibitem{Gall} B. Gallinet and O. J. F. Martin, ``Influence of
electromagnetic interactions on the line shape of plasmonic Fano
resonances". ACS nano \textbf{5}(11):8999 (2011).

\bibitem{Ali} A. Artar, A. Ali Yanik and H. Altug, ``Directional double Fano
resonances in plasmonic hetero-oligomers". Nano Lett. \textbf{11}(9):3694
(2011).

\bibitem{AliM5} S. M. Mozvashi, M. R. Givi, M. B. Tagani, ``The effects of
substrate and stacking in bilayer borophene" Scientific Reports \textbf{12}
(1), 1-10 (2022).

\bibitem{Kamran} K. Ullah, M. T. Naseem , \"{O}. E.M\"{u}stecaplioglu,
``Tunable multiwindow magnomechanically induced transparency, Fano
resonances, and slow-to-fast light conversion". Phys. Rev. A \textbf{102}%
(3): 033721 (2020).

\bibitem{Sabur} A. B. Sabur , A. B. Bhattacherjee. ``Magnomechanically
induced absorption and switching properties in a dispersively coupled
magnon-qubit system". J. Appl. Phys. \textbf{132}, 123104 (2022).

\bibitem{FWM1} L. Shang, B. Chen, L. L. Xing, J. B. Chen, H. B. Xue, and K.
X. Guo, ``Controllable four-wave mixing response in a dual-cavity hybrid
optomechanical system". Chin. Phys. B \textbf{30}(5):295 (2021).

\bibitem{FWM2} M. Maeda, ``General Remarks: Tunable Sources between Vacuum
Ultraviolet and Far Infrared Region". Rev. Laser Eng. \textbf{17}(11):744
(1989).

\bibitem{AliM3} H. Allahverdi, Ali Motazedifard, A. Dalafi, D. Vitali, and
M. H. Naderi, ``Homodyne coherent quantum noise cancellation in a hybrid
optomechanical force sensor" Physical Review A 106 (2), 023107 (2022)

\bibitem{FWM3} J. M. Yuan, W. K. Feng, P. Y. Li, Y. Q. Zhang, and Y. P.
Zhang, ``Controllable vacuum Rabi splitting and optical bistability of
multi-wave-mixing signal inside a ring cavity". Phys. Rev. A \textbf{86}(6)
063820 (2012).

\bibitem{FWM4A} H. J. Chen, H. W. Wu, J. Y. Yang, X. C. Li, Y. J. Sun, Y.
Peng. ``Controllable Optical Bistability and Four-Wave Mixing in a
Photonic-Molecule Optomechanics". Nanoscale Res. Lett. \textbf{14}(1):73
(2019).

\bibitem{FWM5} K. Li, Y. Feng, X. Li, B. Gu, Y. Cai, and Y. Zhang,
``Generation of competitive and coexisting two pairs of biphotons in single
hot atomic vapor cell". Ann. Phys. \textbf{532}(8):2000283 (2020).

\bibitem{FWM6} C. Jiang, Y. Cui, H. Liu. ``Controllable four-wave mixing
based on mechanical vibration in two-mode optomechanical systems". EPL
(Europhysics Letters) \textbf{104}(3):34004 (2013).

\bibitem{FWM7} X. F. Wang, B. Chen. ``Four-wave mixing response in a hybrid
atom-optomechanical system". JOSA B \textbf{36}(2):162 (2019).

\bibitem{Tang} X. Zhang, C. L. Zhang, L. Zou, ``Cavity magnomechanics". Sci.
Adv. \textbf{2} e1501286 (2016).

\bibitem{Heyroth} F. Heyroth, et al., ``Monocrystalline freestanding
three-dimensional yttrium-iron-garnet magnon nanoresonators," Phys. Rev.
Applied. \textbf{12}, 054031 (2019)

\bibitem{Gros} H. Huebl, C. W. Zollitsch, J. Lotze, F. Hocke, A. Marx, R.
Gross, and S. T. B. Goennenwein, ``High cooperativity in coupled microwave
resonator ferrimagnetic insulator hybrids". Phys. Rev. Lett. \textbf{111}%
(12):127003 (2013).

\bibitem{Jiee} J. Li, S. Y. Zhu, and G.S. Agarwal, ``Squeezed states of
magnons and phonons in cavity magnomechanics". Phys. Rev. A \textbf{99}%
(2):021801 (2019).

\bibitem{Usami} D. L. Quirion, Y. Tabuchi, A. Gloppe, K. Usami and Y.
Nakamura, ``Hybrid quantum systems based on magnonics". Appl. Phys. Express
\textbf{12}(7):070101 (2019).

\bibitem{AliM4} Ali Motazedifard, A, Dalafi and M, H, Naderi, ``A Green's
function approach to the linear response of a driven dissipative
optomechanical system" Journal of Physics A: Mathematical and Theoretical
\textbf{54} (21), 215301 (2021)

\bibitem{XLi} Xiyun Li, et. al. ``Phase control of the transmission in
cavity magnomechanical system with magnon driving" J. Appl. Phys. 128,
233101 (2020).

\bibitem{Deng} D. Zhang, X. M. Wang, T. F. Li, X. Q. Luo, W. Wu, F. Nori and
J. You, ``Cavity quantum electrodynamics with ferromagnetic magnons in a
small yttrium-iron-garnet sphere". npj Quantum Inf. \textbf{1}(1):15014
(2015).

\bibitem{Rome} C. Gonzalez-Ballestero, D. Hummer, J. Gieseler, and O.
Romero-Isart, ``Theory of quantum acoustomagnonics and acoustomechanics with
a micromagnet". Phys. Rev. B \textbf{101}(12):125404 (2020).

\bibitem{Cui} C. Kong, B. Wang, Z.-X. Liu, H. Xiong and Y. Wu,
``Magnetically controllable slow light based on magnetostrictive forces".
Opt. express \textbf{27}(4):5544 (2019).


\bibitem{Den} D. Zhang, X. Q. Luo, Y. P. Wang, T.-F. Li and J.Q. You,
``Observation of the exceptional point in cavity magnon-polaritons". Nat.
commun. \textbf{8}(1):1368 (2017).

\bibitem{Tann} H. Tan, ``Genuine photon-magnon-phonon
Einstein-Podolsky-Rosen steerable nonlocality in a continuously-monitored
cavity magnomechanical system". Phys. Rev. Research \textbf{1}(3):033161
(2019).

\bibitem{Asjad} M. Asjad, Jie Li, Shi-Yao Zhu and J. Q. You, ``Magnon
squeezing enhanced ground-state cooling in cavity magnomechanics". arXiv
2022:2203.10767.

\bibitem{Cong} M. S. Ding, L. Zheng and C. Li, ``Phonon laser in a cavity
magnomechanical system". Sci. Rep. \textbf{9}(1):15723 (2019).

\bibitem{Annl} L. Wang, Z. X. Yang, Y. M. Liu, C. H. Bai, D. Y. Wang, S.
Zhang and H. F. Wang, ``Magnon blockade in a PT-symmetric-like cavity
magnomechanical system". Ann. Phys. \textbf{532}(7):2000028 (2020).

\bibitem{CAPE} C. A. Potts, E. Varga, V. Bittencourt, S. V. Kusminskiy, and
J. P. Davis, ``Dynamical backaction magnomechanics" Phys. Rev. X \textbf{11}%
, 031053 (2021).

\bibitem{disp2} J. Li, S. Y. Zhu, G. S. Agarwal., ``Squeezed states of
magnons and phonons in cavity magnomechanics". Phys. Rev. A 99(2):021801
(2019).

\bibitem{mag} J. Li, S. Y. Zhu, G. S. Agarwal., ``Magnon-photon-phonon
entanglement in cavity magnomechanics". Phys. Rev. Lett. \textbf{121}%
(20):203601 (2018).

\bibitem{Kittel} C. Kittel., ``Interaction of spin waves and ultrasonic
waves in ferromagnetic crystals". Phys. Rev. \textbf{110}(4):836 (1958).

\bibitem{ZRCS} Z. Y. Fan , R. C. Shen , Y, P. Wang, J. Li, J. Q. You,
``Optical sensing of magnons via the magnetoelastic displacement". Phys.
Rev. A \textbf{105} 033507 (2022).

\bibitem{Zha} X. Zhang, C. L. Zou, L. Jiang, H. X. Tang., ``Strongly coupled
magnons and cavity microwave photons". Phys. Rev. Lett. \textbf{113}(15),
156401 (2014).

\bibitem{Ballestero} C. Gonzalez-Ballestero, D. H\"{u}mmer, J. Gieseler,
Romero-Isart O. ``Theory of quantum acoustomagnonics and acoustomechanics
with a micromagnet". Phys. Rev. B \textbf{101} (12):125404 (2020).

\bibitem{CGEN} C. Genesa, A. Marib, D. Vitali, P. Tombesi., ``Quantum
Effects in Optomechanical Systems". \textbf{57}:33-86 (2009).

\bibitem{Holstein} T. Holstein, H. Primakoff ``Field dependence of the
intrinsic domain magnetization of a ferromagnet". Phys. Rev. \textbf{58}%
:1098 (1940).

\bibitem{DFW} D. F. Walls, G. J. Milburn, Quantum Optics (Springer), Berlin.
Heidelberg (1994).

\bibitem{AMEPJD} A. Sohail, Y. Zhang, M. Usman, C. S. Yu, ``Controllable
optomechanically induced transparency in coupled optomechanical systems".
Eur. Phys. J. D \textbf{71}(4):103 (2017).

\bibitem{AMPS} A. Sohail, R. Ahmed, R. Zainab, C. S. Yu,: ``Enhanced
entanglement and quantum steering of directly and indirectly coupled modes
in a magnomechanical system". Phys. Scr. \textbf{97}075102 (2022)

\bibitem{ASIJTP} A. Sohail, et. al., ``Magnon-Phonon-Photon Entanglement via
the magnetoelastic Coupling in a Magnomechanical System". IJTP \textbf{61}%
:174 (2022).

\bibitem{AMPS2} A. Sohail, A. Hassan, R. Ahmed, C. S. Yu,, ``Generation of
enhanced entanglement of directly and indirectly coupled modes in a
two-cavity magnomechanical system". Quantum Inf. Process. \textbf{21} 207
(2022)


\bibitem{KQGS} K, Qu, G. S. Agarwal., ``Fano resonances and their control in
optomechanics". Phys. Rev. A \textbf{87}(6) 063813 (2013).

\bibitem{RBWD} R. Becker and W. D\"{o}ring, ``Ferromagnetismus" Springer,
Berlin (1939).

\bibitem{KIT} C. Kittel, ``Physical Theory of Ferromagnetic Domains". Rev.
Mod. Phys. \textbf{21} 541 (1949).

\bibitem{RJW} J. W. Rao, et al, ``Analogue of dynamic Hall effect in cavity
magnon polariton system and coherently controlled logic device, Nat. Commun.
10 2934 (2019).

\bibitem{LGQ} G. Q. Luo , Z. F. Hu, Y. Liang, L. Y. Yu and L. L. Sun ,
``Development of low profile cavity backed crossed slot antennas for planar
integration" IEEE Trans. Antennas Propag. 57 2972 (2009).
\end{thebibliography}
\end{document}